\documentclass{article}
\usepackage{fortschritte}
\usepackage{amsmath}\usepackage{amssymb}
\usepackage{latexsym}\usepackage{cite}

\newcommand{\be}{\begin{equation}}
\newcommand{\ee}{\end{equation}}

\newcommand{\ka}{\kappa}
\newcommand{\mcA}{{\mathcal A}}
\newcommand{\mcD}{{\mathcal D}}
\newcommand{\mcF}{{\mathcal F}}
\newcommand{\mcL}{{\mathcal L}}
\newcommand{\mcN}{{\mathcal N}}
\newcommand{\mcO}{{\mathcal O}}
\newcommand{\mcV}{{\mathcal V}}
\newcommand{\mcW}{{\mathcal W}}

\def\beq{\begin{equation}}                     %
\def\eeq{\end{equation}}                       %
\def\bea{\begin{eqnarray}}                     
\def\eea{\end{eqnarray}}                       
\def\beq{\begin{equation}}
\def\eeq{\end{equation}}
\def\pl{\partial}

\def\al{\alpha}
\def\bt{\beta}

\def\ka{\kappa}
\def\si{\sigma}
\def\Si{\Sigma}
\def\te{\theta}

\def\ep{\epsilon}
\def\ze{\zeta}

\def\l{\left (}
\def\r{\right )}

\def\fr{\frac}
\def\la{\label}

\begin {document}                 
\def\titleline{
%
%
%
The Radion in 5D Conformal SUGRA:\\                        
Superspace Action\footnote{Talk presented at the Ahrenshoop'04 
conference (Berlin, August 23-27, 2004).}
}
%
%
%
%
\def\authors{
%
%
%
%
%
Filipe Paccetti Correia\1ad , Michael G. Schmidt\1ad ,
Zurab Tavartkiladze\2ad                          
}
\def\addresses{
%
%
%
%
\1ad                                        %
 Institut f\"ur Theoretische Physik,
Universit\"at Heidelberg\\ 
Philosophenweg 16, 69120 Heidelberg, Germany
\\
\2ad                                        
 Physics Department, Theory Division, CERN, CH-1211 Geneva 23, Switzerland
   %
}
\def\abstracttext{
%
%
We shortly review our new superfield formalism in the framework of Fujita, 
Kugo, Ohashi 5D conformal supergravity, in particular with an $S^1/Z_2$ 
orbifold.
The radion of the fifth dimension is embedded in two related superfields, a chiral and a general multiplet and is linked to the radion superfield of rigid SUSY. The superspace action of the gauge sector is of the Chern-Simons type. We also present the superspace action for hypermultiplets and discuss the role of compensators.
The presented formalism should be very useful for applications.
We demonstrate this for obtaining the RS solution.

}
\large
\makefront
\section{Introduction}

Extra dimensions and branes, suggested by string theory, offer fascinating new 
aspects for gravity, grand unification, and inflationary cosmology. In 
the 'simple' case of a compactification of one extra dimension on an 
$S^1/Z_2$ orbifold - realized in the Horava-Witten approach to M-Theory 
\cite{hor} - one can discuss AdS 'warped' geometry 
\cite{Randall:1999ee}. Here supersymmetry (SUSY) 
is very useful in order to improve the UV behavior; the SUSY breaking scale 
may be controlled by the inter brane distance. In the case of two types of  
fields, those residing in the full 5D bulk and those based only on the 
branes, we need a consistent treatment of 5D and 4D supergravity (SUGRA). 
Following the Mirabelli-Peskin approach  \cite{mir} for the case of global 
SUSY one  preferably should have an off shell SUGRA formulation; i.e. with 
auxiliary fields not integrated out, in order to formulate correct brane-bulk 
interactions. This was pioneered in the work of Zucker \cite{zucker}. It is 
based on tensor calculus, the gauge algebra really closes, and Lagrangians can 
be constructed after the formulation of symmetry breaking/fixation. 
However in this approach one still has to undergo redefinitions, 
there is preferably a tensor compensator and the final effective 4D SUGRA is 
not of the minimal type. Still it is almost exclusively used in most 
phenomenologically oriented work \cite{phenSUGRA}. 
In several steps of improvements the subsequent work of Fujita, 
Kugo and Ohashi (FKO) \cite{Fujita:2001kv} led to a very 
efficient 5D superconformal tensor calculus, 
includes gauge fields, and leads to the old Poincare SUGRA on the 4D 
boundaries 
after strict conformal gauge fixing. 
In their papers this was worked out with component fields.
In our work \cite{PaccettiCorreia:2004ri}, we have attempted to present this elegantly in terms 
of 4D superfields, with the prospect to model applications 
\cite{Correia:2004pz}, \cite{prep} close to the known 
proposals in the case of global SUSY \cite{5Dsusy} - \cite{heb} where superfields are very helpful.

\section{Reduction of 5D Supermultiplets to 4D Superfields}
A particular question goes for the 'radion' superfield introduced
(mostly in the rigid SUSY case) in the literature \cite{Marti:2001iw}
\beq
T=(R+{\rm i}A_y,~\psi_{y-},~F_T)~,
\la{Tsup}
\eeq
with $\pi R$ the (fluctuating) inter brane distance, $A_y$ the $5^{\rm th}$
component of the graviphoton, $\psi_{y-}$ the corresponding gravitino and
$F_T$ an auxiliary field with a SUSY breaking VEV. Can it be embedded into 
SUGRA at all and if yes, in which form? The answer will be negative in a 
strict sense but the notion of a radion multiplet is still useful in some 
approximation. It is quite helpful
to start with a look to the rigid 5D SUSY case and to the reduction of its 
gauge and hyper - multiplets to 4D superfields \cite{5Dsusy} - \cite{heb}. 
Skipping that we here immediately go to the 5D conformal SUGRA of FKO: 
The gauge 
fields of the superconformal group  reside in the Weyl multiplet
\be
                      (e_{\mu}^{\,a},~b_{\mu},~\psi_{\mu}^{\,i},~V_{\mu}^{\,ij},~v_{ab},~\chi^i,~D),
\la{weylmul}
\ee
where the first four elements correspond to the Lorentz transformations, 
dilatations, supersymmetric transformations, and the $SU(2)_R$, and the rest 
are auxiliary fields closing the algebra off shell. Furthermore there are 
vector gauge supermultiplets ${\mathbb V}^I$, $I=1, 2, \cdots $
(nonabelian in general)
\be
{\mathbb V}^I=(M,~\Omega^i,~W_{\mu},~Y^{ij})^I~,   
\ee
and hypermultiplets
\beq
{\mathbb H}^{\al }=({\cal A}^{\al }_i,~\ze^{\al }, {\cal F}^{\al }_i)~.
\la{hypmult}
\eeq
with indices $i=1,2$ of $SU(2)_R$ and $\al =1, 2, \cdots , 2r$.

Also compensator fields are required: a vector multiplet $V^0$
[the 'graviphoton', (note the change in FKO from $1^{\rm st}$ to the second paper in citation \cite{Fujita:2001kv})
which is needed for fixing the dilatation and special (super)conformal gauge 
transformations] and hypermultiplets ${\mathbb H}^{\al }$, usually a pair, fixing 
$SU(2)_R$. The most important structure in the following will be a 
norm function $ \mcN (M)$ and 'prepotential' $P$
\be
                      \mcN (M)=\kappa c_{IJK} M^I M^J M^K=-2P(M)
\la{normprep}
\ee 
in terms of which the former gauge fixing conditions read 
\be
                       \mcN (M)=\kappa^{-2},\quad \mcN_I(M)\Omega^{I}=0,\quad {\hat\mcD}_{a}\mcN (M)=0,
\ee
Integrating out the $D$-field of (\ref{weylmul}) one obtains  $\mcA^2=-2\mcN$
which requires the compensator hypermultiplet(s) and breaks $SU(2)_R$. The 
component action based on these multiplets is given by FKO. It is split into 
a supergravity (gauge) part, a Chern-Simons part, a hypermultiplet part and an 
additional 
${\cal L}_{\rm aux}$-part where the auxiliary fields show up such that 
${\cal L}_{\rm aux}$ gives their values on shell. In table 1 we give a 
typical list of the field content, already split up according to the $Z_2$ 
orbifold parity of the fields.

\begin{table}[!h]
\begin{center}
\begin{tabular}{|c|c|}
\hline 
$Z_2$ parity & Field content\\
\hline\hline
\multicolumn{2}{|c|}{Weyl multiplet}\\
\hline
+ & $e^a_{\mu},~e^5_y,~\psi_{\mu+},~\psi_{y-},~b_{\mu},~V_{\mu}^3,~V_y^{1,2},~v^{5a},~\chi_+,~D $\\
\hline
-& $e^5_{\mu},~e^a_y,~\psi_{\mu-},~\psi_{y+},~b_{y},~V_{y}^3,~V_{\mu}^{1,2},~v^{ab},~\chi_-$ \\
\hline\hline
\multicolumn{2}{|c|}{Vector multiplet}\\
\hline
$\Pi_V$ & $M,~W_y,~\Omega_-,~Y^{1,2}$ \\
\hline
$-\Pi_V$ & $W_{\mu},~\Omega_+,~Y^3$ \\
\hline\hline
\multicolumn{2}{|c|}{Hypermultiplet}\\
\hline
$\Pi_{\hat\alpha}$ & $\mcA^{2{\hat\alpha}-1}_1,~\mcA^{2{\hat\alpha}}_2,~\zeta^{\hat\alpha}_-,~\mcF^{2{\hat\alpha}-1}_1,~\mcF^{2{\hat\alpha}}_2$\\
\hline
$-\Pi_{\hat\alpha}$ & $\mcA^{2{\hat\alpha}-1}_2,~\mcA^{2{\hat\alpha}}_1,~\zeta^{\hat\alpha}_+,~\mcF^{2{\hat\alpha}-1}_2,~\mcF^{2{\hat\alpha}}_1$\\
\hline
\end{tabular}
\end{center}
\caption{Field Content and $Z_2$ Parities}
\end{table}

The 5D vector multiplet in FKO notation decomposes into a 4D vector multiplet 
\be
           V^I=(W_{\bar\mu},2\Omega_+,2Y^{3}-{\hat {\mathcal D}}_5 M)^I,       \la{redV4d}
\ee
where $\Omega_+\equiv \Omega_R^1+\Omega_L^2$        
and the covariant derivative of $M$ is
${\hat {\mathcal D}}_{5} M=(\partial_{5}-b_{5})M-ig[W_{5},M]-2\ka i{\bar \psi}_{5}\Omega $,
(with $b_5$ set to zero at the end)
and a 4D chiral multiplet
$\Sigma^I=(\phi^I,\chi^I_R,F^I_{\phi})$, with
\be
           \begin{array}{ccl} \phi^I & = & \frac{1}{2}(e^5_y M^I-iW_y^I),\\
                              \chi^I & = & 2e^5_y\gamma_5\Omega_{-}^I-2i\ka\psi_{y-}M^I,\\
                              F^I_{\phi} & = & -e^5_y(Y^1+iY^2)^I-i\kappa M^I(V^1_y+iV^2_y)+i\ka{\bar\psi}_{y-}(1+\gamma_5)\Omega_{-}^I, \end{array}         
\la{redSi4d}
\ee
Composing now 
$T=2\ka {\Sigma}_T\equiv 2(\mcN_I/3\mcN)\Sigma^I$,
and using the gauge fixing conditions above, for the components 
$(T_{\rm sc}, {\chi}_T,  {F}_{T} )$ of $T$ we obtain
\be
           \begin{array}{rcl} T_{\rm sc} & = & e^5_y -i\ka {B}_y~,\\
                              {\chi}_T & = & -i4\ka \psi_{y-},\\
                              {F}_{T} & = & 2\ka e^5_y({t}^1+i{t}^2)-i2\ka (V^1_y+iV^2_y). \end{array}         
\ee
Notice, that the real part of the scalar component $T_{\rm sc}$ is the radion.
Therefore, $T$ may be called 'radion supermultiplet' even though only in the $\ka \to 0$ limit it really becomes a supermultiplet.
Expanding (\ref{normprep}) around $M^0$ we have
\be
         \frac{{\mathcal N}_I}{ 3\kappa{\mathcal N}}=
(c_{000})^{1/3}\delta^0_I+\mcO (\ka M^{I\neq 0}).    
\ee
$V^I$ and $\Si^I$ have opposite $Z_2$ parities. Thus one of them has a zero 
mode and unsuppressed interactions on the branes. Following FKO, there is still 
another (general type) 4D supermultiplet
\be\label{eq:V_5def}
          \mcV_5^I=(~M,-2i\gamma_5\Omega_-,~2Y^1,~2Y^2,~{\hat F}_{a5}+2\ka v_{a5}M,~\lambda^{\mcV_5},~D^{\mcV_5})^I,
\ee
where $\lambda^{\mcV_5}$ and $D^{\mcV_5}$ is an auxiliary field component 
given in ref. \cite{Fujita:2001kv}.
{}For $\ka \to 0$ it reduces to
\be
                \mcV_5^I|_{\ka=0}=(e^5_y)^{-1}(\Sigma+\Sigma^+)^I-\partial_5 V^I.    
\ee
and thus in the case $I=0$ is related to the radion. For $\ka \neq 0$
$e_y^5$ has to be enlarged to an (even) real 4D supermultiplet
$W_y$ containing fragments of the 5D Weyl multiplet
\be
             {\mathbb W}_y=(~e^5_y,-2\ka\psi_{y-},-2\ka V_y^2,~2\ka V_y^1,-2\ka v_{ay},~\lambda^{{\mathbb W}_y},~D^{{\mathbb W}_y}),
\ee  
with
\be
             \lambda^{{\mathbb W}_y}=\tfrac{i}{4}\ka e^5_y\gamma_5\chi_+ +2\phi_{y+}+2\ka^2\gamma_5\gamma^bv_{b5}\psi_{y-},
\ee
\be
             D^{{\mathbb W}_y}=\ka^2e^5_y\left[\tfrac{1}{4}D-(v_{a5})^2\right]-2f_y^5+\tfrac{i}{4}\ka^2{\bar\chi}_+ \gamma_5\psi_{y-}~. 
\ee
In this case
\be\label{eq:V_5defsuper}
               \mcV_5= \frac{\Sigma+\Sigma^+ -\partial_y V}{{\mathbb W}_y} +\cdots   
\ee
On the orbifold the 5D supersymmetry transformations are split into even and 
odd ones. The first correspond just to the usual $N=1$ SUSY, the odd SUSY 
transformation appears on the branes only in $\pl_5$-derivative combinations.
It is remarkable that this ${\cal V}_5$ does {\it not} transform under the
'odd SUSY'.

It is not difficult to recognize that there is some overlap between
$\fr{1}{2}(T+T^{\dagger })$ and ${\mathbb W}_y$. Namely
\beq
{\mathbb W}_y=\fr{1}{2}(T+T^{\dagger })+\cdots ~.
\la{overl}
\eeq
Relations (\ref{eq:V_5defsuper}), (\ref{overl}) will be used for deriving the radion interaction with gauge and hyper-multiplets.

Now we turn to the hypermultiplet in (\ref{hypmult}). It can be split as
${\mathbb H}^{\al }=({\mathbb H}^{2\hat{\al }-1}, {\mathbb H}^{2\hat{\al }})$
with $\hat{\al }=1, 2, \cdots , r$.
The scalar components satisfy
\beq
({\cal A}_2^{2\hat{\al }})^*={\cal A}_1^{2\hat{\al }-1}~,~~~~
({\cal A}_1^{2\hat{\al }})^*=-{\cal A}_2^{2\hat{\al }-1}~,
\la{relcond}
\eeq
(and similarly for ${\cal F}$ components). These constraints clearly relate 
${\mathbb H}^{2\hat{\al }-1}$ with ${\mathbb H}^{2\hat{\al }}$.
Therefore, for each 
$\hat{\al }$ only 
four real scalar components are independent\footnote{For a more detailed 
discussion about hypermultiplets see \cite{Fujita:2001kv}.}.
For a given $\hat{\al}$, the 5D hypermultiplet decomposes into a 
pair of $\mcN=1$ 4D chiral superfields
with opposite orbifold parities \cite{Fujita:2001kv}:
$$
H=\l {\cal A}^{2\hat{\al }}_2,~-2{\rm i}\ze^{2\hat{\al }}_R,~
({\rm i}M_*{\cal A}+{\hat{\mcD}}_5{\cal A})^{2\hat{\al }}_1\r, 
$$
\beq
H^c=\l {\cal A}^{2\hat{\al }-1}_2=-({\cal A}^{2\hat{\al }}_1)^*,~
-2{\rm i}\ze^{2\hat{\al }-1}_R,~
({\rm i}M_*{\cal A}+{\hat {\mcD}}_5{\cal A})^{2\hat{\al }-1}_1\r, 
\la{hyperdec}
\eeq
with 
$$
M_*{\cal A}^{\al }_i=igM^I(t_I)^{\al }_{\bt }{\cal A}^{\bt }_i+
{\cal F}^{\al }_i~,
$$
\beq
\hat{\mcD}_{5 }{\cal A}^{\al }_i=
\pl_{5 }{\cal A}^{\al }_i-i
gW_{5 \bt }^{\al }{\cal A}^{\bt }_i-
W_{5 }^0\fr{1}{\al }{\cal F}^{\al }_i-
\ka V_{5 ij}{\cal A}^{\al j}-
2\ka {\rm i}\bar \psi_{5 i}\ze^{\al }~,
\la{covders}
\eeq 
where $(t_I)^{\al }_{\bt }$ is the generator of the gauge group
$G_I$

In the following we will use the notation
\beq
{\bf H}\equiv ({\bf H}_1,~{\bf H}_2)=(H, ~H^c)~.
\la{hyper}
\eeq
If we have more than one 5D hypermultiplet($r>1$), the index $\hat{\al}$ should be present, ${\bf H}^{\hat{\al}}\equiv ({\bf H}_1,~{\bf H}_2)^{\hat{\al}}=(H, ~H^c)^{\hat{\al}}$.

\section{Superspace Action and Application for Warped Solution}

The invariant superfields presented above can now be used to construct the 
5D Lagrangian. 
The vector part Lagrangian, in terms of superfields, turns out to be
\be\begin{split}\label{eq:vecsuperlagr}
          e_{(4)}^{-1}{\mathcal L}_5= & \frac{1}{4}\int d^2\theta\left(P_{IJ}(2\Sigma){\mathcal W}^{I\alpha}{\mathcal W}_{\alpha}^J-\frac{1}{6}P_{IJK}{\bar D}^2 (V^I D^{\alpha}\partial_y V^J-D^{\alpha}V^I\partial_y V^J){\mathcal W}_{\alpha}^K\right)\\
                  &\qquad {}+\textup{h.c.}+\int d^4 \theta\, {\mathbb W}_y \,2P({\mathcal V}_5).
\end{split}\ee
This superspace Lagrangian gives the (non gravitational) vector Lagrangian of 
FKO, including couplings to the radion sector.
With the Weyl weight $w(d^n\theta)=n/2$, one sees that the r.h.s. of the above expression has Weyl weight four, which indeed compensates for the transformation properties of $e_{(4)}\equiv\textup{det}\,{e_{(4)}}_{\mu}^a$ under dilatations since $w(e_{(4)})=-4$. 
${\mathbb W}_y$ and $\Sigma$ transform in a non-trivial way 
\cite{Fujita:2001kv} under the \emph{odd} 5D superconformal transformations, unlike, for instance, what happens with ${\mathcal V}_5$. To compensate for this non-trivial behaviour one may need to add terms to ${\mathbb W}_y$, which include derivatives of the corresponding gauge fields, build out of those components of the 5D Weyl multiplet which are odd under orbifold parity.

The superspace action and relations given above allow to get the interaction 
of the radion superfield with gauge supermultiplets. For demonstrative 
purposes together with a compensator gauge supermultiplet ${\mathbb V}^{I=0}$ consider one abelian 
gauge supermultiplet ${\mathbb V}^{I=1}$. With the norm function 
$ \ka^{-1}\mcN={M^{0}}^3-M^0 {M^1}^2-\gamma{M^1}^3$ 
from (\ref{eq:V_5defsuper}),  (\ref{overl}),  (\ref{eq:vecsuperlagr})
we obtain in the $\ka \to 0$ limit
\be              2{\mathbb W}_y
P(\mcV_5)\simeq-\kappa^{-2}\frac{T+T^+}{2}+2\frac{(\Sigma^1+\Sigma^{1+}-\partial_y V^1)^2}{T+T^+}+4\kappa\gamma\frac{(\Sigma^1+\Sigma^{1+}-\partial_y V^1)^3}{(T+T^+)^2}.   \ee
These terms  coincide with those given before in the literature 
\cite{Marti:2001iw}, \cite{luty}.

We now turn to a superspace action for the hypermultiplet 
$({\bf H}_1,~{\bf H}_2)$
interacting with a 5D gauge multiplet $(V, ~\Si)$. 
For this purpose we introduce
\beq 
{\bf V}^{ab}=gV\vec q \cdot \vec \si^{ab}~,~~~~~
{\bf \Si}^{ab}=g\Si\vec q \cdot \vec \si^{ab}~,~~~
{\rm with}~~|\vec q|=1~.
\la{gaugeR}
\eeq
This notation turns out to be convenient for constructing the action
invariant under different orbifold parity prescriptions for the vector 
superfields. In the component off-shell formulation it has been already
used in order to gauge the $U(1)_R$ symmetry 
\cite{Fujita:2001kv}.
Here we use it for a general Abelian $U(1)$ gauge symmetry.
%
%

In the case of one $r$-hypermultiplet and one gauge field the hypermultiplet Lagrangian  has the form
\beq
e_{(4)}^{-1}{\cal L}({\bf H})=\int d^4\te\, {\mathbb W}_y 2{\bf H}^{\dagger }_a
(e^{-{\bf V}})^{ab}{\bf H}_b+
\int d^2\te ({\bf H}\ep )_a\l \hat{\pl_y}-
{\bf \Si } \r^{ab}{\bf H}_b
+{\rm h.c.} 
\la{hypAct}
\eeq
where the superoperator $\hat{\pl }_y$ is obtained by promoting ${\pl }_y$
to an operator containing odd (under orbifold parity) elements of the 5D
Weyl multiplet (see a detailed discussion in ref. 
\cite{PaccettiCorreia:2004ri}).

Also compensator hypermultiplets can couple to the 
gauge fields $({\bf V},~{\bf \Si })$, see \cite{Fujita:2001kv}. Since they have negative kinetic terms, for case of compensators $(e^{-{\bf V}})^{ab}$ should simply be replaced by $-(e^{-{\bf V}})^{ab}$.
As usual, the exponent of the first term  in eq.(\ref{hypAct}) 
completes  4D derivatives promoting them to 
derivatives covariant under the gauge transformations. In the same way, an additional coupling to the 4D \emph{Weyl} supermultiplet should be included, in order to covariantize the 4D derivatives of the hypermultiplets with respect to the superconformal symmetries. This can however be bypassed by using the (4D) 
superconformal-invariant D and F-term action formulas 
(see discussions in ref. \cite{PaccettiCorreia:2004ri}). The superconformal covariant derivatives in the fifth direction, which appear in the kinetic terms, arise in part from the F-term coupling $[({\bf H}\ep )_a (\hat{\pl_y}-{\bf \Si }^{ab}){\bf H}_b]_F$. While ${\bf \Si }^{ab}$ takes care of the gauge invariance, $\hat{\pl_y}$ induces some of the pieces of the superconformal covariant derivatives of $\mcA$ and $\zeta$. The remaining terms are due to the coupling of ${\mathbb W}_y$ to the hypermultiplets in the D-term coupling $[{\mathbb W}_y{\bf H}^{\dagger }{\bf H}]_D$. 
 
One can check that all the relevant (non gravitational)  
couplings of FKO \cite{Fujita:2001kv} involving hypermultiplets 
are reproduced by expression (\ref{hypAct}).
The first term in (\ref{hypAct}) gives the radion interaction with the hypermultiplets:
$[(T+T^{\dagger }){\bf H}^{\dagger }{\bf H}]_D$. One more comment is in order:
The supermultiplet ${\mathbb V}^{I=0}=(V_0, \Si_0)$ gauges the central charge symmetry $U(1)_Z$. This is a compensating gauge supermultiplet and as discussed above is closely related to the radion. On the other hand, in the covariant derivatives of eq.  (\ref{covders}), $I=0$ does not participate in
$M^I(t_I)^{\al }_{\bt }$ and $W^{\al }_{5\bt }$ but acts only on an auxiliary component ${\cal F}^{\al }_i$. Because of this, $\Si_0$ (i.e. the radion superfield) does not appear in the second term of the action (\ref{hypAct}).

As an application of the superspace action presented above, we derive the RS solution. We consider the case of a single compensator charged under $5D$ gauge multiplet
${\mathbb V}\equiv V_I{\mathbb V}^I$ (the combination of multiplets
${\mathbb V}^I=(\Sigma^I,~V^I)$) with negative orbifold parity and an odd
gauge coupling 
$G(y)=g\epsilon (y)$ (i.e. we take in (\ref{gaugeR}) $q_{1,2}=0$). Then 
for the compensator hypermultiplet we have
$$
             e_{(4)}^{-1}{\cal L}=
-\int d^4\te\,{\mathbb W}_y 2\l H^{\dagger }e^{-G (y) V}H+H^{c\dagger }e^{G (y) V}H^c\r +
$$
\be\label{eq:warpedlagran}
\int d^2\te (H^c\hat{\pl_y}H-H\hat{\pl_y}H^c - 2G (y)H^c\Si H)+{\rm h.c.}.
\ee
We are considering the metric 
\be
                 ds^2=e^{2\sigma(y)}dx_{\mu}dx^{\mu}-(e^5_y)^2dy^2,
\ee
and use the following Weyl rescaling for the states involved:
$$
 e^a_{\mu}\to e^{-\sigma}e^a_{\mu}~,~~~
H=e^{3\sigma/2}\kappa^{-1}-\theta^2 e^{5\sigma/2}F^{*},\quad
H^c=-\theta^2 e^{5\sigma/2}F^{c*}~,
$$
\be
              \Sigma=\left(\tfrac{1}{2}(e^5_y M-iA_y),\cdots\right),\quad V=e^{\sigma}\left(A_{\mu},~e^{\sigma/2}2\Omega_+,~e^{\sigma} D\right), \quad {\mathbb W}_y=(e^{-\sigma}e^5_y,\cdots).
\ee
Integration of all F and D-terms leads finally to the action 
\cite{PaccettiCorreia:2004ri}
\be\begin{split}\label{eq:RSgeneral1}
              e^{-1}\mcL_{H+V}\supset & -\frac{1}{4}\mcN_{IJ}\partial_a M^I\partial^a M^J+ 6M^3_5 (\partial_5\sigma)^2\\
                                      &\hspace{48pt} +M_5^3 \left(\frac{1}{2}g^2 M^2 + M_5^3 g^2 V_I V_J \mcN^{IJ} + e^y_5 (\partial_y \epsilon (y))gM \right).
\end{split}\ee
Parameterizing the \emph{very special} manifold defined by 
$\mcN(M)=\kappa^{-2}$  with fields 
$\phi^i$,
$g_{ij}(\phi)\equiv -\frac{1}{2}\mcN_{IJ}\frac{\partial M^I}{\partial\phi^i}\frac{\partial M^J}{\partial\phi^j},\quad \mcW(\phi) \equiv M_5^3 g V_I M^I(\phi)$,
eq. (\ref{eq:RSgeneral1}) can be rewritten as
\be\label{eq:warplagrfin}
         e^{-1}\mcL=\frac{1}{2} g_{ij}(\phi)\partial_a\phi^i\partial^a\phi^j + 6M^3_5 (\partial_5\sigma)^2 -V_B(\phi)+2e^y_5 [\delta(y)-\delta(y-\pi R)]\,\mcW(\phi),   
\ee
where the bulk potential is
$V_B(\phi)=\frac{1}{2}g^{ij}\mcW_i\mcW_j-\frac{2}{3M_5^3}\mcW^2$.
Without physical (bulk) vector multiplets 
 $\mcW_i=0$ we get for brane tensions $\tau=\pm 2\mcW$ 
(with $\tau=\pm \sqrt{-6M^3_5 V_B}$) the automatic tuning with a bulk cosmological constant as in the RS model.
{}The equation for the warp-factor $\sigma(y)$ follows from 
eq.\eqref{eq:warplagrfin}, giving the solution 
$\sigma(y)=\pm\sqrt{-V_B/6M^3_5}|y|$.

Concluding, let us note that the presented formalism is very efficient to 
study e.g. the gauging of $R$-symmetries and derive supersymmetric warped 
solutions also in the presence of FI terms
\cite{Correia:2004pz}. Also it turns out to be quite useful and economical 
for studying some type of inflationary scenarios in five dimensions 
\cite{prep}.


\end{document}